\documentclass[aps,prl,showpacs,twocolumn,groupedaddress,amssymb]{revtex4}
\usepackage{graphicx}% Include figure files
\usepackage{dcolumn}% Align table columns on decimal point
\usepackage{bm}% bold math

\begin{document}

\title{Hard X-ray or Gamma Ray Source Based on the two-stream instability and Backward Raman Scattering}% \\

\author{S. Son}
\affiliation{169 Snowden Lane, Princeton, NJ, 08540}
%\author{Sung Joon Moon}
%\affiliation{PACM, Princeton University, Princeton, NJ 08544}
%\author{J.~Y. Park}
%\affiliation{Los Alamos National Laboratory}

\begin{abstract}

A new scheme of hard x-ray or gamma ray light source is considered. 
The excitation of the Langmuir wave in an ultra dense electron beam
via the two-stream instabilities 
and the interaction of the excited Langmuir wave with the visible light-laser
results in hard x-ray or gamma ray via three-wave interaction. 
  The analysis suggests that the hard x-ray with the wave-length  as small as 0.03 nm can be achieved. 
The plausible parameter for the practical use is proposed  and 
the comparison with the conventional methods are provided.
\end{abstract}

\maketitle

Since its initial discovery by Wilhelm Rontgen, 
the x-ray has been utilized in many applications and its numerous light sources have been invented~\cite{sonttera, colson, songamma,Gallardo, sonltera,freelaser, freelaser2, Free, Free2}. 
With the great advances in other scientific areas,
there have been exponentionally growing interests for \textit{intense} x-ray light applications.
 %there have been exponentionally growing interests for \textit{intense} soft x-ray light, 
%as it could create new industries in the atomic spectroscopy~\cite{soft, soft2, soft3}, the dynamical imaging of fast biological processes~\cite{bio, bio2} and the next-generation semi-conductor lithography~\cite{litho, litho2, litho3, litho4}.
An intense  hard x-ray or gamma ray source would open up 
many possible commercial applications,  including   the atomic spectroscopy~\cite{soft, soft2, soft3}, the dynamical imaging of fast biological processes~\cite{bio, bio2} and the next-generation semi-conductor lithography~\cite{litho, litho2, litho3, litho4}.
Yet, it is very hard to achieve intense hard x-ray or gamma ray source~\cite{laser,laser1,gain,sonband,freelaser, freelaser2, Free, Free2}.

Noting the recent progress in the dense electron beam source~\cite{monoelectron, ebeam}
 and the intense visible-light lasers~\cite{cpa, cpa2, cpa4}, 
the author proposes  a new scheme of a hard x-ray or gamma ray laser  
based on the two-stream instability and the backward Raman scattering (BRS). 
%The excitation of the Langmuir waves via the 
% two-stream instability inside a dense relativistic electron beam 
% and the backward Raman scattering of the excited Langmuir waves with the visible-light laser could result in the hard x-ray or gamma ray. %
%The main idea can be summarized as follows. 
If  two electron beams propagate 
in the same direction with the different drift velocities, 
the Langmuir waves could be excited inside the electron beams via the two-stream instability, and 
  the BRS of the excited Langmuir waves 
with an counter-propagating visible-light laser could result in the hard x-ray or gamma ray. 
%In the previous researches of the x-ray generation based on the BRS, 
% there have been no proposed method of generating an appropriate Langmuir wave (or seed x-ray) needed for the BRS. 
The current work has two major merits.  
First, 
the seed x-ray is very hard to generate but pre-requisite for the x-ray generation via the BRS, and  the technological options to create an intense seed are severely limited so far.
If the Langmuir waves (or the seed x-ray) are generated  via the two-stream instability,  there is no need to create  an intense seed.
  % In any prior art,  no attempt has been made to 
%generate an appropriate Langmuir wave (or seed x-ray) needed for the BRS.
Second,
as the Langmuir waves  are directly excited by the two-stream instability,   not  by the ponderomotive interaction between a seed x-ray and an intense laser,
the required intensity of a visible light laser is considerably lower.  
%utilizing 
%There has been a recent attempt to generate  Langmuir waves inside an electron beam for the various purposes~\cite{nicolai}. 
%The research in this paper extends furhter focused on dense electron beams and hard x-ray generaton. 
% deals with the similar idea but with focuse on an dense electron beams and 
The analysis suggests that 
the hard x-ray up to the wave length of 0.03 nm is possible
and the requirement of the laser intensity is moderate. 
The regime of the practical interest is identified 
and the advantage (disadvantage) over the straight use of the BRS~\cite{brs, brs2, brs3, drake} are discussed. 
%Very intense hard x-ray or gamma ray might be possible 
%using moderately intense laser beams. 
%The energy conversion efficiency from the visible-light laser to the hard x-ray is estimated and the optimal physical parameter for practical use are suggested.%

\begin{figure}
\scalebox{0.3}{
\includegraphics[width=1.7\columnwidth, angle=270]{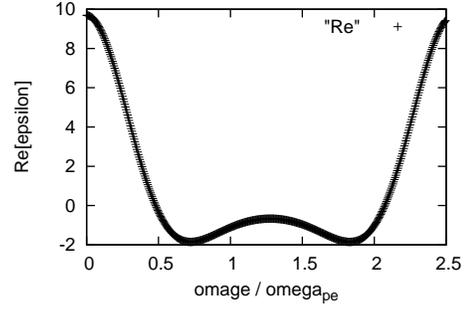}}
\caption{\label{fig1}
The real part of the dielectric function $\epsilon $ as a function of the frequency for the classical plasmas. The x-axis is $\omega / \omega_{pe} $ and the y-axis is $Re[\epsilon] $.  
In this example, $n_0 = 5 \times 10^{23} / \mathrm{cc} $, $\gamma_0 = 200$, 
$T_e = 200 \  \mathrm{eV} $, $v_0 /c = 0.15 $, $k \lambda_{de} = 0.34 $ 
so that  $k v_0 /\omega_{pe} \cong = 2.55 $.  
The local maxima of the real part at $ \omega = 1.3  \ \omega_{pe} $ is less than 0, which  is the threshold condition of the two-stream instability. 
}
\end{figure}

\begin{figure}
\scalebox{0.3}{
\includegraphics[width=1.7\columnwidth, angle=270]{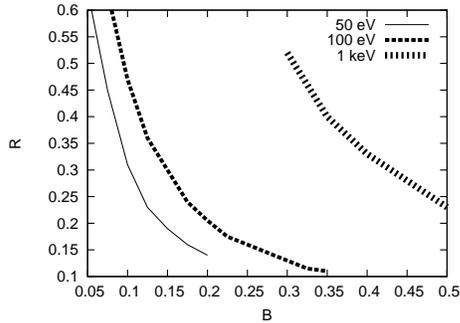}}
\caption{\label{fig2}
The threshold wave vector $k_C $ as a function of the drift velocity $v_0$. 
In this example, $n_0 = 5 \times 10^{23} / \mathrm{cc} $ and  $\gamma_0 = 200$. 
The y-axis is  $ \mathrm{R} = k_C \lambda_{de} $ and the y-axis is $\mathrm{B} = v_0 / c$. 
Three cases when $T_e = 50 \  \mathrm{eV,} \ 200 \ \mathrm{eV,} \ 1000 \ \mathrm{eV}$ are considered. 
There is no two-stream instability when 
 $v_0 / c < 0.06 $ for $T_e = 50  \ \mathrm{eV}$, 
 $v_0 / c < 0.12 $ for $T_e = 200 \  \mathrm{eV}$ and 
 $v_0 / c < 0.3 $ for $T_e = 1000 \  \mathrm{eV}$
}
\end{figure}

To begin with, 
Consider two dense relativistic electron beams with the relativistic factor $\gamma_0 < \gamma_1$, where $\gamma_0 = (1-\beta_1^2)^{-1/2}$ ($\gamma_1 = (1-\beta_2^2)^{-1/2}$), $\beta_1 = v_1/c$ ($\beta_2 = v_2/c$) and $v_1 $ ($v_2$) is the velocity of the electron beam 0 (1). 
 For simplicity, the electron density of two beams are assume to be the same $n_0$. 
Due to the initiative from the fast igniter concept of the 
inertial confinement fusion~\cite{monoelectron, ebeam, tabak,sonprl,sonpla, sonpla2, sonpla3, sonchain}, 
the electron density as high as  $n_0 \cong 5\times 10^{23} \  / \mathrm{cc}$
are being contemplated,  to which 
the electron density is assumed to be  comparable.  
%let us consider  two co-propagating relativistic electron beams. 
It is most convenient to use the co-moving frame with the electron beams. 
In this frame, one of the electron beam is stationary and 
the other beam has the drift velocity 
%the drift of the other beam is also assumed to be non-relativistic ($v_0 < c $). 
given  as $v_0 = (v_1 - v_2) / (1 - \beta_1 \beta_2 ) $.
%where $v_1$ ($v_2$) is the beam velocity in the laboratory frame and 
%  $\beta_1 = v_1 / c$ ($\beta_2=v_2/c$). 
%Denote the electron temperature of the beams as $T_e$, which we assume to be the same for the both beams. 
In the co-moving frame, the both beams have the density 
$n_e \cong n_0 /\gamma_0$. 
%Our assumption is that the well prepared pair of the electron beams will excite
%the suitable Langmuir wave for the backward Raman scattering. 
%Let us now consider a counter-propagating visible-light laser. 
An appropriate Langmuir wave excited by the two-stream instability 
and  the visible-light laser could  emit
the x-ray (seed pulse) in the beam direction via the BRS. 
%This visible light laser might excite the hard x-ray via the backward Raman scattering with the excited Langmuir waves. 
Conventionally,  the visible-light laser (the emitted hard x-ray) is called as
the pump laser (seed laser).
Denoting the wave vector of the visible-light laser in the laboratory frame as
$k_0$, it will be derived later that the appropriate Langmuir wave for the BRS has the wave vector    $k_3 \cong 4 \gamma_0 k_{p0}$. 
Then, the higher the $k_3$ is, the higher the frequency of the x-ray can be emitted; it is important to estimate 
%For a given electron temperature $T_e$  the possible maximum wave vector of the Langmuir wave is given as $k_{max} \cong 1/\lambda_{de} $, where 
%$\lambda_{de}^2 = T_e \gamma_0 / 4 \pi n_0 e^2 $ is the Debye length
%of the electron beam in the co-moving frame. 
the possible highest wave vector of the Langmuir wave excitable  by  the two-stream instability. 
%If the beams has a energy spread $\delta E/E$, the electron temperature in the co-moving  frame is comparable to $T_e \cong (\delta E/E)^2 m_e c^2$ from the fact that  $\delta v/c \cong \delta E / E$, where $\delta v $ is the velocity spread of the beam in the co-moving frame. 
%If the energy spread is given as $ 0.01 < \delta E /E < 0.1$, the 
%elec tron temperature in the co-moving frame would be $ 25 \ \mathrm{eV} < T_e < 2.5 \ \mathrm{keV} $. 
%The analysis of the two-stream instability can be  performed using the conventional dielectric function approach.
The criteria of the two-stream instability would be that 1), for a fixed wave-vector $k$,  there is the local maxima of the dielectric function $\epsilon$   as a function of the wave frequency $\omega$
and 2), the value of the local maxima is less than zero.
The longitudinal dielectric function of a plasma 
is given as   
 
\begin{equation} 
\epsilon(\mathbf{k}, \omega) = 1 + \frac{4 \pi e^2 }{k^2} \Sigma \chi_i \mathrm{.}
\end{equation} 
where the summation is over the group of particle species and $  \chi_i $ is the particle susceptibility. 
In classical plasmas, the susceptibility is given as 

\begin{equation}
 \chi_i^C(k, \omega) = \frac{n_iZ_i^2}{m_i} \int \left[ \frac{ \mathbf{k} \cdot \mathbf{\nabla}_v f_i }{\omega - \mathbf{k} \cdot \mathbf{v} }\right]d^3 \mathbf{v} 
\end{equation} 
where $m_i$ ($Z_i$, $n_i$) is the particle mass (charge, density) and $f_i $ is the distribution with the normalization $\int f_i d^3 \mathbf{v} = 1$.  
For the case of our two group of electrons, it is given as 
\begin{equation} 
\epsilon =  1 + (4 \pi e^2/k^2) ( \chi_e^C(\omega, k) +  \chi_e^C(\omega-\mathbf{k}\cdot \mathbf{v}_0, k) \mathrm{.}
 \label{eq:ele}
\end{equation} 
One example, where the Langmuir wave  is susceptible 
to the two-stream instability, is  shown in Fig~(\ref{fig1}).
In Fig.~(\ref{fig2}), we draw the threshold wave vector $k_c$, over which 
all the plasma Langmuir waves become stable to the two-stream instability. 
In the figure, we plot the maximum $k_C(v_0)$ as a function of the drift velocity $v_0$ for three different electron temperature. 
Such analysis suggests that 
the optimal regime is characterized by 
$2.5 < k v_0 \sqrt{\gamma_0} / \omega_{pe}<3.5$ and $k \lambda_{de} < 0.5$. 
 In general, the most prominent condition  is  

\begin{equation} 
 k_3 \lambda_{de} \leq 0.5 \label{eq:tt} \mathrm{.} 
\end{equation}
For a fixed electron temperature and the  electron density, there is a lower bound of the drift velocity $v_0/c$ under which 
the plasma is stable to the two-stream instability for any wave vector, 
which is also illustrated in Fig.~(\ref{fig2}).

Eq.~(\ref{eq:tt}) suggests that  the lower electron temperature the plasma has, the higher Langmuir wave vector 
can be excited via the two-stream instability.
However, the electron temperature is limited by the energy spread of 
the electron beams, which cannot be lowered to a certain degree. 
If the beam has the energy spread $\delta E/ E$ in the laboratory frame, 
the electron temperature in the co-moving frame is given as $T_e \cong (\delta E/E)^2 m_e c^2 $ from the fact that $\delta E/E \cong \delta v/c $. 
If the energy spread is $0.01 < \delta E / E < 0.1$, 
then the electron temperature would be $25 \ \mathrm{eV} < T_e < 2.5 \ \mathrm{keV} $.
As shown in the figure, for the  electron temperature of 200 eV, 
the maximum wave vector is given as $k_c \cong 0.5 / \lambda_{de} $ when 
$v_0 / c \cong 0.15$. 

If $c k_3 /\omega_3 \cong \gamma_0^{3/2} (ck_{p0} / \omega_{pe} ) \gg 1$,
 then  $k_3 \cong 4 \gamma_0 k_{p0} $ and $\omega_3 \cong \omega_{pe} / \sqrt{\gamma_0} $. 
The condition given in Eq.~(\ref{eq:tt}) can be rewritten as 
\begin{equation} 
\lambda_{de}^L < \frac{1}{k_{s0}} \times \left( \frac{k_{s0} }{k_{p0}} \right) \mathrm{,} \label{eq:cond}
\end{equation}
where $\lambda_{de}^L = \sqrt{T_e/ 4 \pi n_0 e^2} $, $k_{s0} $ is the wave vector of the x-ray and $k_{p0}$ is the wave vector of the visible-light laser. 
For a fixed $k_{s0} $, the lower the $k_{p0} $ is, the higher the electron temperature can be.  For this reason, 
the infra-red laser with the wave length of $10  \ \mu \mathrm{m} $ to $20 \ \mu \mathrm{m} $ 
is more advanageous than ND:YAG laser with  the wave length of $1  \ \mu \mathrm{m} $

With an appropriate Langmuir wave excited by the two-stream instability, 
the 1-D BRS three-wave interaction 
in the \textit{co-moving} frame between the pump,  
the seed  and a Langmuir wave  
is described  by~\cite{McKinstrie}:
\begin{eqnarray}
\left( \frac{\partial }{\partial t} + v_p \frac{\partial}{\partial x} + \nu_1\right)A_p  = -ic_p A_s A_3  \nonumber \mathrm{,}\\
\left( \frac{\partial }{\partial t} + v_s \frac{\partial}{\partial x} + \nu_2\right)A_s  = -ic_s A_p A^*_3   \label{eq:2} \mathrm{,} \\
\left( \frac{\partial }{\partial t} + v_3 \frac{\partial}{\partial x} + \nu_3\right)A_3  = -ic_3 A_p A^*_s  
\nonumber \mathrm{,}
\end{eqnarray}
where $A_i= eE_{i1}/m_e\omega_{i1}c$  is 
the ratio of  the electron quiver velocity of the pump pulse ($i=p$)
and the seed pulse ($i=s$)  relative to the velocity of the light $c$,
 $E_{i1}$ is the electric field of the E\&M pulse, 
 $A_3 = \delta n_1/n_1$ is the the Langmuir wave amplitude,
$\nu_1 $ ($\nu_2$) is the rate of the inverse bremsstrahlung  
of the pump (seed), 
$\nu_3$ is the plasmon decay rate, 
$ c_i = \omega_3^2/ 2 \omega_{i1}$ for $i=p, s$, $c_3 = (ck_3)^2/2\omega_3$, 
$\omega_{s}$ ($\omega_{p}$) is the wave frequency of the x-ray (the pump laser) and  $\omega_{3} \cong \omega_{pe} / \sqrt{\gamma_0} $  is  the plasmon  wave frequency.  In the co-moving frame with the electron beam,   
%the electron density decreases to  
%$n_1 = n_0 / \gamma_0$ due to the length dilation and 
the wave vector of the light wave satisfies 
the usual dispersion relationship, 
$\omega_1^2 =  2  \omega_{pe}^2/\gamma_0+ c^2 k_1^2$, where 
$\omega_1$ ($k_1$) is the wave frequency (vector). 
Denote  the wave vector  (the corresponding wave frequency) of 
the pump laser (the seed pulse or soft x-ray) in the co-moving frame 
as  $k_{p1}$, $k_{s1} $, $\omega_{p1} $ and   $\omega_{s1} $, 
 and  the laboratory-frame counterparts  
 as  $k_{p0}$, $k_{s0} $,  $\omega_{p0} $ and $\omega_{s0} $.
The Lorentz transform prescribes the following relationship: 
\begin{eqnarray} 
\omega_{p0} &=& \gamma_0 \left[ \sqrt{2\omega_{pe}^2/\gamma_0 + c^2 k_{p1}^2 } - vk_{p1} \right] \mathrm{,}  \label{eq:lorentz1} \\  \nonumber \\
k_{p0} &=&  \gamma_0 \left[ k_{p1} - \frac{\omega_{p1} }{c}  \frac{v_0}{c} \right] \mathrm{,} \label{eq:lorentz2} \\ \nonumber \\ 
 \omega_{s0} &=& \gamma_0 \left[ \sqrt{2\omega_{pe}^2/\gamma_0 + c^2 k_{s1}^2 } + 
vk_{s1} \right] \mathrm{,}  \label{eq:lorentz3} \\  \nonumber \\ 
k_{s0} &=&  \gamma_0 \left[ k_{s1} + \frac{\omega_{s1} }{c}  \frac{v_0}{c} \right]\mathrm{.} \label{eq:lorentz4} \\ \nonumber 
\end{eqnarray}
Using Eqs.~(\ref{eq:lorentz1}), (\ref{eq:lorentz2}), (\ref{eq:lorentz3}) and  (\ref{eq:lorentz4}), 
the pump laser (seed pulse or soft x-ray) can be transformed 
from the co-moving frame to the laboratory frame or vice versa. 
%I analyze the BRS in the co-moving  frame and 
%transform the results  to the laboratory frame. 
The energy and momentum conservation of Eq.~(\ref{eq:2}) leads to  
\begin{eqnarray} 
 \omega_{p1} &=& \omega_{s1} + \omega_{3}  \nonumber 
\mathrm{,} \nonumber \\  
k_{p1} &=& k_{s1} +k_3\mathrm{,} \label{eq:cons}  
\end{eqnarray}
where $k_3$ is the plasmon wave vector.
For a given pump frequency $\omega_{p0} $, 
$k_{p1} $ ($\omega_{p1} $) is  obtained from  Eq.~(\ref{eq:lorentz1}), 
  $k_{s1} $ ($\omega_{s1}$) is  from  Eq.~(\ref{eq:cons}) and    
$k_{s0} $ ($\omega_{s0}$) is  from 
 Eqs.~(\ref{eq:lorentz3}) and (\ref{eq:lorentz4}). 
In the limit when $ck_{s1} \gg \omega_3 $,  
$\omega_{s0} \cong 2 \gamma_0 (\omega_{p1} - \omega_3)$ or  

\begin{equation} 
\omega_{s0} \cong 4\gamma_0^2 \left[ \omega_{p0} - 2\sqrt{2} \omega_{pe} (\gamma_0)^{-3/2}\right] \mathrm{,}\label{eq:down}
\end{equation} 
using  $\omega_{p1} \cong 2 \gamma_0  \omega_{p0}$ and 
$\omega_3 \cong \omega_{pe} /\sqrt{\gamma_0} $.
The equation \ref{eq:down} describes   the frequency up-shift of the pump pulse into the hard x-ray 
 by the relativistic Doppler's effect. 

The first one  in  Eq.~(\ref{eq:2}) is the  most relevant for us 
 as the excited Langmuir wave is given by $A_3= \delta n_e /n_e$.  
 The mean-free path of the laser to the BRS is estimated to be 
\begin{equation} 
l_b \cong c (2 \sqrt{\omega_{s1}\omega_{p1}}  /\omega_3^2 ) (1/A_3) 
\mathrm{.} \label{eq:mean}
\end{equation}
The mean-free-path from 
the Thomson scattering (the Compton scattering) is  $ l_t \cong 1/n\sigma_t $ with $\sigma_t = (mc^2 / e^2)^2 $.  For an example, when $n_1 \cong 5\times 10^{23} / \mathrm{cc} $,   
 $l_t \cong 0.2 \ \mathrm{cm} $ and $l_b \cong (10^{-6} / A_3) ( 2 \omega_{s1}  /\omega_3) \ \mathrm{cm}$.  Even for $A_3 \cong 0.001 $, the hard x-ray  radiation by the BRS is considerably stronger than the Thomson scattering or $l_t \gg l_b $.

As the first example, consider the electron beams with $n_0  =  5\times 10^{23} \ / \mathrm{cc} $, $\gamma_0 = 200 $ and $T_e = 25 \ \mathrm{eV} $. 
For the visible-light laser with $\lambda = 10 \ \mu\mathrm{m}$, 
 $k_3 \lambda_{de}  = 0.37$. 
The Langmuir wave for the BRS will be unstable when $0.06 < v_0 /c < 0.08 $ as shown in Fig.~(\ref{fig2}). 
For the visible-light laser with 
$\lambda = 20 \ \mu\mathrm{m}$,  $k_3 \lambda_{de} = 0.18 $. 
the Langmuir wave for the BRS will be unstable when $0.12 < v_0 / c < 0.16 $. 
The emitted light will have the wave length of 
 0.06 nm (0.12 nm) for $\lambda = 10 \ \mu\mathrm{m}$
($\lambda = 20 \ \mu\mathrm{m}$).
As the second example, consider the same beam but with $\gamma_0 = 100 $ and $T_e = 200 \ \mathrm{eV} $.
For the visible-light laser with $\lambda = 10 \ \mu\mathrm{m}$ ($\lambda = 20 \ \mu\mathrm{m}$), 
$k_3 \lambda_{de}  = 0.37$ ($k_3 \lambda_{de}  = 0.18$).
The Langmuir wave for the BRS is unstable
 for the visible-light laser with
 when $0.1 < v_0 / c < 0.15$ ($0.1 < v_0 / c < 0.3$).
The emitted light will be the wave length of 0.25 nm (0.5 nm) for $\lambda = 10 \ \mu\mathrm{m}$
 ($\lambda = 20 \ \mu\mathrm{m}$).
As the third example, consider the same electron beam with $T_e = 1 \ \mathrm{keV} $. 
For the visible-light laser with  $\lambda = 10 \ \mu\mathrm{m}$ ($\lambda = 20 \ \mu\mathrm{m}$), $k_3 \lambda_{de} = 0.83 $ ( $k_3 \lambda_{de} = 0.43 $)
and the plasma Langmuir waves is stable to the two-stream instability
for  $\lambda = 20 \ \mu\mathrm{m}$ when  $0.3 < v_0 / c < 0.35$.
The emitted light will be the wave length of (0.5 nm) for $\lambda = 20 \ \mu\mathrm{m}$

In summary, 
we propose a new scheme of a gamma ray or hard x-ray based on 
the two-stream instability and the backward Raman scattering. 
The excitation of  Langmuir waves in an ultra dense relativistic electron beam via the two-stream instabilities and 
the subsequent interaction of the excited Langmuir waves with the infra-red laser via the backward Raman scattering results in 
the hard x-ray and gamma ray.  
With the highest electron  density possible with the current technologies, 
the gamma ray or hard x-ray in the range of $ 1 \ \mathrm{keV} < \hbar \omega < 50  \ \mathrm{keV}$ is possible. 

In comparison to the previous schemes~\cite{sonbrs, brs, brs2, brs3, drake},  the current sheme has  some disadvantages and  advantages. 
One disadvantage is the requirement 
of the low-energy spread of the electron beams; 
two electron beams with very low-energy spread are needed
in order to excite the Langmuir wave with the highest possible wave vector. 
There are a few advantages.
First, the required intensity of the  visible-light laser is lowered considerably in the current scheme.
  As discussed in Eq.~(\ref{eq:cond}), 
the required laser intensity is too high for 
the infra-red laser under the previous scheme but achievable under the current scheme.   
%,   as discussed in Eq.~(\ref{eq:cond}), 
%both in the new scheme and in the direct BRS.
%the required laser intensity cannot be . 
%while the required intensity is moderate in the new scheme. 
%while the intensity requirement of the current scheme is moderate. 
Second, 
the previous  methods need  an uniform electron beam 
but the current scheme has much relaxed requirement of the uniformity. 
Third, 
%another prominent advantage of this scheme over the direct BRS is the inverse bremsstrahlung.
%For dense electron beams, the plasma can heat up very fast due to the inverse bremsstrahlung especially when the pump pulse has the high intensity.
%This heating is detrimental since the Langmuir wave cannot be excited no longer due to the rising temperature.  In the current scheme, 
%the moderately intense laser beams are needed not like the direct BRS method, 
%where the intense pump laser is required. 
%one clear advantage of the current scheme over the
%$direct BRS ~\cite{sonbrs, brs, brs2, brs3, drake} 
 the inverse bremsstrahlung is not a concern in the current scheme.  
%The rate of the inverse bremsstrahlung could be very high in dense plasmas. 
In the regime of our interest ($n_e = 10^{23} - 10^{24}  \ / \mathrm{cc} $),
 the inverse bremsstrahlung rate is  as high as
 $0.01 \times  \omega_{pe} $
and   the heating by the inverse bremsstrahlung increases 
the electron temperature in a few hundred Langmuir periods if the pump laser is very intense as in the direct BRS~\cite{sonbackward}.
%the heating by the inverse bremsstrahlung increases 
%the electron temperature in a few hundred Langmuir periods~\cite{sonbackward} and , 
After the heating from the inverse bremsstrahlung,
 the appropriate Langmuir wave could be no longer excited due to the  Landau damping. 
But, in the current scheme, the intensity of the pump laser can be very low. 
so that 
the inverse bremsstrahlung is less of concern. 
%The excitation of the Langmuir wave is done by the two-stream instability 
%instead of the intense visible-light laser. 
%the by-products of the Langmuir wave excitation.
%In other words,  the cost of the exciting the Langmuir wave can be 
%cheaper in our scheme than the direct BRS scheme. 

% Those pros and cons should be analyzed carefully in order to access the practical applications of the proposed scheme.  
In this paper, it is assumed that 
the Langmuir wave for the BRS will be excited as long as it it unstable to the 
two-stream instability.
However, 
the future work needs to check through the simulation and the theoretical analysis how intense the excited Langmuir wave is. 
The full adequacy of the current scheme can be further  validated by the study along this line, which is beyond the scope of this paper.

\bibliography{tera2}% Produces the bibliography via BibTeX.

\begin{thebibliography}{40}
\expandafter\ifx\csname natexlab\endcsname\relax\def\natexlab#1{#1}\fi
\expandafter\ifx\csname bibnamefont\endcsname\relax
  \def\bibnamefont#1{#1}\fi
\expandafter\ifx\csname bibfnamefont\endcsname\relax
  \def\bibfnamefont#1{#1}\fi
\expandafter\ifx\csname citenamefont\endcsname\relax
  \def\citenamefont#1{#1}\fi
\expandafter\ifx\csname url\endcsname\relax
  \def\url#1{\texttt{#1}}\fi
\expandafter\ifx\csname urlprefix\endcsname\relax\def\urlprefix{URL }\fi
\providecommand{\bibinfo}[2]{#2}
\providecommand{\eprint}[2][]{\url{#2}}

\bibitem[{\citenamefont{Son et~al.}(2012{\natexlab{a}})\citenamefont{Son, Moon,
  and Park}}]{sonttera}
\bibinfo{author}{\bibfnamefont{S.}~\bibnamefont{Son}},
  \bibinfo{author}{\bibfnamefont{S.~J.} \bibnamefont{Moon}}, \bibnamefont{and}
  \bibinfo{author}{\bibfnamefont{J.~Y.} \bibnamefont{Park}},
  \bibinfo{journal}{Optics Letters} \textbf{\bibinfo{volume}{37}},
  \bibinfo{pages}{5172} (\bibinfo{year}{2012}{\natexlab{a}}).

\bibitem[{\citenamefont{Colson}(1985)}]{colson}
\bibinfo{author}{\bibfnamefont{W.~B.} \bibnamefont{Colson}},
  \bibinfo{journal}{Nucl.~Inst.~Meth.~Phys. A} \textbf{\bibinfo{volume}{237}},
  \bibinfo{pages}{1} (\bibinfo{year}{1985}).

\bibitem[{\citenamefont{Son and Moon}(2012)}]{songamma}
\bibinfo{author}{\bibfnamefont{S.}~\bibnamefont{Son}} \bibnamefont{and}
  \bibinfo{author}{\bibfnamefont{S.~J.} \bibnamefont{Moon}},
  \bibinfo{journal}{Phys.~Plasmas} \textbf{\bibinfo{volume}{19}},
  \bibinfo{pages}{063102} (\bibinfo{year}{2012}).

\bibitem[{\citenamefont{Gallardo}(1988)}]{Gallardo}
\bibinfo{author}{\bibfnamefont{J.~C.} \bibnamefont{Gallardo}},
  \bibinfo{journal}{IEEE J.~Quantum.~Elec.} \textbf{\bibinfo{volume}{24}},
  \bibinfo{pages}{1557} (\bibinfo{year}{1988}).

\bibitem[{\citenamefont{Son et~al.}(2012{\natexlab{b}})\citenamefont{Son, Moon,
  and Park}}]{sonltera}
\bibinfo{author}{\bibfnamefont{S.}~\bibnamefont{Son}},
  \bibinfo{author}{\bibfnamefont{S.~J.} \bibnamefont{Moon}}, \bibnamefont{and}
  \bibinfo{author}{\bibfnamefont{J.~Y.} \bibnamefont{Park}},
  \bibinfo{journal}{Phys.~Plasmas} \textbf{\bibinfo{volume}{19}},
  \bibinfo{pages}{114503} (\bibinfo{year}{2012}{\natexlab{b}}).

\bibitem[{\citenamefont{Carr et~al.}(2002)\citenamefont{Carr, Martin, Mckinney,
  Jordan, Neil, and Williams}}]{freelaser}
\bibinfo{author}{\bibfnamefont{G.~L.} \bibnamefont{Carr}},
  \bibinfo{author}{\bibfnamefont{M.~C.} \bibnamefont{Martin}},
  \bibinfo{author}{\bibfnamefont{W.~R.} \bibnamefont{Mckinney}},
  \bibinfo{author}{\bibfnamefont{K.}~\bibnamefont{Jordan}},
  \bibinfo{author}{\bibfnamefont{G.~R.} \bibnamefont{Neil}}, \bibnamefont{and}
  \bibinfo{author}{\bibfnamefont{G.~P.} \bibnamefont{Williams}},
  \bibinfo{journal}{Nature} \textbf{\bibinfo{volume}{420}},
  \bibinfo{pages}{153} (\bibinfo{year}{2002}).

\bibitem[{\citenamefont{Williams}(2002)}]{freelaser2}
\bibinfo{author}{\bibfnamefont{G.~P.} \bibnamefont{Williams}},
  \bibinfo{journal}{Review of Scientific Instruments}
  \textbf{\bibinfo{volume}{73}}, \bibinfo{pages}{1461} (\bibinfo{year}{2002}).

\bibitem[{\citenamefont{Wabnitz et~al.}(2002)\citenamefont{Wabnitz, Bittner,
  de~Castro, Dohrmann, Gurtler, Laarmann, Laasch, Schulz, Swiderski, von
  Haeften et~al.}}]{Free}
\bibinfo{author}{\bibfnamefont{H.}~\bibnamefont{Wabnitz}},
  \bibinfo{author}{\bibfnamefont{L.}~\bibnamefont{Bittner}},
  \bibinfo{author}{\bibfnamefont{A.~R.~B.} \bibnamefont{de~Castro}},
  \bibinfo{author}{\bibfnamefont{R.}~\bibnamefont{Dohrmann}},
  \bibinfo{author}{\bibfnamefont{P.}~\bibnamefont{Gurtler}},
  \bibinfo{author}{\bibfnamefont{T.}~\bibnamefont{Laarmann}},
  \bibinfo{author}{\bibfnamefont{W.}~\bibnamefont{Laasch}},
  \bibinfo{author}{\bibfnamefont{J.}~\bibnamefont{Schulz}},
  \bibinfo{author}{\bibfnamefont{A.}~\bibnamefont{Swiderski}},
  \bibinfo{author}{\bibfnamefont{K.}~\bibnamefont{von Haeften}},
  \bibnamefont{et~al.}, \bibinfo{journal}{Nature}
  \textbf{\bibinfo{volume}{420}}, \bibinfo{pages}{482} (\bibinfo{year}{2002}).

\bibitem[{\citenamefont{Emma et~al.}(2004)\citenamefont{Emma, Bane, Cornacchia,
  Huang, Schlarb, Stupakov, and Walz}}]{Free2}
\bibinfo{author}{\bibfnamefont{P.}~\bibnamefont{Emma}},
  \bibinfo{author}{\bibfnamefont{K.}~\bibnamefont{Bane}},
  \bibinfo{author}{\bibfnamefont{M.}~\bibnamefont{Cornacchia}},
  \bibinfo{author}{\bibfnamefont{Z.}~\bibnamefont{Huang}},
  \bibinfo{author}{\bibfnamefont{H.}~\bibnamefont{Schlarb}},
  \bibinfo{author}{\bibfnamefont{G.}~\bibnamefont{Stupakov}}, \bibnamefont{and}
  \bibinfo{author}{\bibfnamefont{D.}~\bibnamefont{Walz}},
  \bibinfo{journal}{Phys. Rev. Lett} \textbf{\bibinfo{volume}{92}},
  \bibinfo{pages}{074801} (\bibinfo{year}{2004}).

\bibitem[{\citenamefont{Yin et~al.}(1997)\citenamefont{Yin, Kasrai, Fuller,
  Bancroft, Fyfe, and Tan}}]{soft}
\bibinfo{author}{\bibfnamefont{Z.}~\bibnamefont{Yin}},
  \bibinfo{author}{\bibfnamefont{M.}~\bibnamefont{Kasrai}},
  \bibinfo{author}{\bibfnamefont{M.}~\bibnamefont{Fuller}},
  \bibinfo{author}{\bibfnamefont{G.~M.} \bibnamefont{Bancroft}},
  \bibinfo{author}{\bibfnamefont{K.}~\bibnamefont{Fyfe}}, \bibnamefont{and}
  \bibinfo{author}{\bibfnamefont{K.~H.} \bibnamefont{Tan}},
  \bibinfo{journal}{Wear} \textbf{\bibinfo{volume}{202}}, \bibinfo{pages}{`72}
  (\bibinfo{year}{1997}).

\bibitem[{\citenamefont{Shapiro et~al.}(2005)\citenamefont{Shapiro, Thibault,
  Beetz, Elser, Howells, Jacobsen, Kirz, Lima, Miao, Neiman et~al.}}]{soft2}
\bibinfo{author}{\bibfnamefont{D.}~\bibnamefont{Shapiro}},
  \bibinfo{author}{\bibfnamefont{P.}~\bibnamefont{Thibault}},
  \bibinfo{author}{\bibfnamefont{T.}~\bibnamefont{Beetz}},
  \bibinfo{author}{\bibfnamefont{V.}~\bibnamefont{Elser}},
  \bibinfo{author}{\bibfnamefont{M.}~\bibnamefont{Howells}},
  \bibinfo{author}{\bibfnamefont{C.}~\bibnamefont{Jacobsen}},
  \bibinfo{author}{\bibfnamefont{J.}~\bibnamefont{Kirz}},
  \bibinfo{author}{\bibfnamefont{E.}~\bibnamefont{Lima}},
  \bibinfo{author}{\bibfnamefont{H.}~\bibnamefont{Miao}},
  \bibinfo{author}{\bibfnamefont{A.~M.} \bibnamefont{Neiman}},
  \bibnamefont{et~al.}, \bibinfo{journal}{Proc.~Nat.~Acad.~Sci.}
  \textbf{\bibinfo{volume}{102}}, \bibinfo{pages}{15343}
  (\bibinfo{year}{2005}).

\bibitem[{\citenamefont{Workman et~al.}(1997)\citenamefont{Workman, Nantel,
  Maksimchuk, and Umstadter}}]{soft3}
\bibinfo{author}{\bibfnamefont{J.}~\bibnamefont{Workman}},
  \bibinfo{author}{\bibfnamefont{M.}~\bibnamefont{Nantel}},
  \bibinfo{author}{\bibfnamefont{A.}~\bibnamefont{Maksimchuk}},
  \bibnamefont{and}
  \bibinfo{author}{\bibfnamefont{D.}~\bibnamefont{Umstadter}},
  \bibinfo{journal}{Appl.~Phys.~Lett.} \textbf{\bibinfo{volume}{70}},
  \bibinfo{pages}{312} (\bibinfo{year}{1997}).

\bibitem[{\citenamefont{Muchmore et~al.}(2012)\citenamefont{Muchmore, Sattler,
  Liang, Meadows, Harlan, Yoon, Nettesheim, Chang, Thompson, Wong
  et~al.}}]{bio}
\bibinfo{author}{\bibfnamefont{S.~W.} \bibnamefont{Muchmore}},
  \bibinfo{author}{\bibfnamefont{M.}~\bibnamefont{Sattler}},
  \bibinfo{author}{\bibfnamefont{H.}~\bibnamefont{Liang}},
  \bibinfo{author}{\bibfnamefont{R.~P.} \bibnamefont{Meadows}},
  \bibinfo{author}{\bibfnamefont{J.~E.} \bibnamefont{Harlan}},
  \bibinfo{author}{\bibfnamefont{H.~S.} \bibnamefont{Yoon}},
  \bibinfo{author}{\bibfnamefont{D.}~\bibnamefont{Nettesheim}},
  \bibinfo{author}{\bibfnamefont{B.~S.} \bibnamefont{Chang}},
  \bibinfo{author}{\bibfnamefont{C.~B.} \bibnamefont{Thompson}},
  \bibinfo{author}{\bibfnamefont{S.~L.} \bibnamefont{Wong}},
  \bibnamefont{et~al.}, \bibinfo{journal}{Nature}
  \textbf{\bibinfo{volume}{381}}, \bibinfo{pages}{335} (\bibinfo{year}{2012}).

\bibitem[{\citenamefont{Whitesides et~al.}(2001)\citenamefont{Whitesides,
  Ostuni, Takayama, X.Jiang, and Ingber}}]{bio2}
\bibinfo{author}{\bibfnamefont{G.~W.} \bibnamefont{Whitesides}},
  \bibinfo{author}{\bibfnamefont{E.}~\bibnamefont{Ostuni}},
  \bibinfo{author}{\bibfnamefont{S.}~\bibnamefont{Takayama}},
  \bibinfo{author}{\bibnamefont{X.Jiang}}, \bibnamefont{and}
  \bibinfo{author}{\bibfnamefont{D.~E.} \bibnamefont{Ingber}},
  \bibinfo{journal}{Annual Review of Biomedical Engineering}
  \textbf{\bibinfo{volume}{3}}, \bibinfo{pages}{335} (\bibinfo{year}{2001}).

\bibitem[{\citenamefont{Kinoshita et~al.}(1989)\citenamefont{Kinoshita,
  Kurihara, Ishii, and Torii}}]{litho}
\bibinfo{author}{\bibfnamefont{H.}~\bibnamefont{Kinoshita}},
  \bibinfo{author}{\bibfnamefont{K.}~\bibnamefont{Kurihara}},
  \bibinfo{author}{\bibfnamefont{Y.}~\bibnamefont{Ishii}}, \bibnamefont{and}
  \bibinfo{author}{\bibfnamefont{Y.}~\bibnamefont{Torii}},
  \bibinfo{journal}{J.~Vac.~Sci.~Technol.~B} \textbf{\bibinfo{volume}{7}},
  \bibinfo{pages}{1648} (\bibinfo{year}{1989}).

\bibitem[{\citenamefont{Ehrfeld and Lehr}(1995)}]{litho2}
\bibinfo{author}{\bibfnamefont{W.}~\bibnamefont{Ehrfeld}} \bibnamefont{and}
  \bibinfo{author}{\bibfnamefont{H.}~\bibnamefont{Lehr}},
  \bibinfo{journal}{Radiation Physics and Chemistry}
  \textbf{\bibinfo{volume}{45}}, \bibinfo{pages}{349} (\bibinfo{year}{1995}).

\bibitem[{\citenamefont{Ito and Okazaki}(2000)}]{litho3}
\bibinfo{author}{\bibfnamefont{T.}~\bibnamefont{Ito}} \bibnamefont{and}
  \bibinfo{author}{\bibfnamefont{S.}~\bibnamefont{Okazaki}},
  \bibinfo{journal}{Nature} \textbf{\bibinfo{volume}{406}},
  \bibinfo{pages}{1027} (\bibinfo{year}{2000}).

\bibitem[{\citenamefont{Gywn et~al.}(1998)\citenamefont{Gywn, Stulen, Sweeney,
  and Attwood}}]{litho4}
\bibinfo{author}{\bibfnamefont{C.~W.} \bibnamefont{Gywn}},
  \bibinfo{author}{\bibfnamefont{R.}~\bibnamefont{Stulen}},
  \bibinfo{author}{\bibfnamefont{D.}~\bibnamefont{Sweeney}}, \bibnamefont{and}
  \bibinfo{author}{\bibfnamefont{D.}~\bibnamefont{Attwood}},
  \bibinfo{journal}{J.~Vac.~Sci.~Technol.~B} \textbf{\bibinfo{volume}{16}},
  \bibinfo{pages}{3142} (\bibinfo{year}{1998}).

\bibitem[{\citenamefont{Afana'ev and Kalynov}(1989)}]{laser}
\bibinfo{author}{\bibfnamefont{Y.}~\bibnamefont{Afana'ev}} \bibnamefont{and}
  \bibinfo{author}{\bibfnamefont{V.~S. Y.~K.} \bibnamefont{Kalynov}},
  \bibinfo{journal}{Sov. J. Quantum Electron} \textbf{\bibinfo{volume}{19}},
  \bibinfo{pages}{1506} (\bibinfo{year}{1989}).

\bibitem[{\citenamefont{Nilsen}(1997)}]{laser1}
\bibinfo{author}{\bibfnamefont{J.}~\bibnamefont{Nilsen}},
  \bibinfo{journal}{J.~Opt.~Soc.~Am.~B} \textbf{\bibinfo{volume}{14}},
  \bibinfo{pages}{1511} (\bibinfo{year}{1997}).

\bibitem[{\citenamefont{Dunn et~al.}(2000)\citenamefont{Dunn, Li, Osterheld,
  Nelson, Hunter, and Shlyaptsev}}]{gain}
\bibinfo{author}{\bibfnamefont{J.}~\bibnamefont{Dunn}},
  \bibinfo{author}{\bibfnamefont{Y.}~\bibnamefont{Li}},
  \bibinfo{author}{\bibfnamefont{A.~L.} \bibnamefont{Osterheld}},
  \bibinfo{author}{\bibfnamefont{J.}~\bibnamefont{Nelson}},
  \bibinfo{author}{\bibfnamefont{J.~R.} \bibnamefont{Hunter}},
  \bibnamefont{and} \bibinfo{author}{\bibfnamefont{V.~N.}
  \bibnamefont{Shlyaptsev}}, \bibinfo{journal}{Phys.~Rev.~Lett.}
  \textbf{\bibinfo{volume}{84}}, \bibinfo{pages}{4834} (\bibinfo{year}{2000}).

\bibitem[{\citenamefont{Kue et~al.}(2010)\citenamefont{Kue, Son, and
  Moon}}]{sonband}
\bibinfo{author}{\bibfnamefont{S.}~\bibnamefont{Kue}},
  \bibinfo{author}{\bibfnamefont{S.}~\bibnamefont{Son}}, \bibnamefont{and}
  \bibinfo{author}{\bibfnamefont{S.~J.} \bibnamefont{Moon}},
  \bibinfo{journal}{Phys.~Plasmas} \textbf{\bibinfo{volume}{17}},
  \bibinfo{pages}{052702} (\bibinfo{year}{2010}).

\bibitem[{\citenamefont{Mangles et~al.}(2004)\citenamefont{Mangles, Murphy,
  Najmudin, Thomas, Collier, Dangor, Divall, Foster, Gallacher, Hooker
  et~al.}}]{monoelectron}
\bibinfo{author}{\bibfnamefont{S.~P.~D.} \bibnamefont{Mangles}},
  \bibinfo{author}{\bibfnamefont{C.~D.} \bibnamefont{Murphy}},
  \bibinfo{author}{\bibfnamefont{Z.}~\bibnamefont{Najmudin}},
  \bibinfo{author}{\bibfnamefont{A.~G.~R.} \bibnamefont{Thomas}},
  \bibinfo{author}{\bibfnamefont{J.~L.} \bibnamefont{Collier}},
  \bibinfo{author}{\bibfnamefont{A.~E.} \bibnamefont{Dangor}},
  \bibinfo{author}{\bibfnamefont{E.~J.} \bibnamefont{Divall}},
  \bibinfo{author}{\bibfnamefont{P.~S.} \bibnamefont{Foster}},
  \bibinfo{author}{\bibfnamefont{J.~G.} \bibnamefont{Gallacher}},
  \bibinfo{author}{\bibfnamefont{C.~J.} \bibnamefont{Hooker}},
  \bibnamefont{et~al.}, \bibinfo{journal}{Nature}
  \textbf{\bibinfo{volume}{431}}, \bibinfo{pages}{535} (\bibinfo{year}{2004}).

\bibitem[{\citenamefont{Tatarakis et~al.}(2003)\citenamefont{Tatarakis, Beg,
  Clark, Dangor, Edwards, Evans, Goldsack, Ledingham, Norreys, Sinclair
  et~al.}}]{ebeam}
\bibinfo{author}{\bibfnamefont{M.}~\bibnamefont{Tatarakis}},
  \bibinfo{author}{\bibfnamefont{F.~N.} \bibnamefont{Beg}},
  \bibinfo{author}{\bibfnamefont{E.~L.} \bibnamefont{Clark}},
  \bibinfo{author}{\bibfnamefont{A.~E.} \bibnamefont{Dangor}},
  \bibinfo{author}{\bibfnamefont{R.~D.} \bibnamefont{Edwards}},
  \bibinfo{author}{\bibfnamefont{R.~G.} \bibnamefont{Evans}},
  \bibinfo{author}{\bibfnamefont{T.~J.} \bibnamefont{Goldsack}},
  \bibinfo{author}{\bibfnamefont{K.~W.~D.} \bibnamefont{Ledingham}},
  \bibinfo{author}{\bibfnamefont{P.~A.} \bibnamefont{Norreys}},
  \bibinfo{author}{\bibfnamefont{M.~A.} \bibnamefont{Sinclair}},
  \bibnamefont{et~al.}, \bibinfo{journal}{Phys.~Rev.~Lett.}
  \textbf{\bibinfo{volume}{90}}, \bibinfo{pages}{175001}
  (\bibinfo{year}{2003}).

\bibitem[{\citenamefont{Strickland and Mourou}(1985)}]{cpa}
\bibinfo{author}{\bibfnamefont{D.}~\bibnamefont{Strickland}} \bibnamefont{and}
  \bibinfo{author}{\bibfnamefont{G.}~\bibnamefont{Mourou}},
  \bibinfo{journal}{Phys.~Fluids} \textbf{\bibinfo{volume}{55}},
  \bibinfo{pages}{447} (\bibinfo{year}{1985}).

\bibitem[{\citenamefont{Perry and Mourou}(1994)}]{cpa2}
\bibinfo{author}{\bibfnamefont{M.~D.} \bibnamefont{Perry}} \bibnamefont{and}
  \bibinfo{author}{\bibfnamefont{G.}~\bibnamefont{Mourou}},
  \bibinfo{journal}{Science} \textbf{\bibinfo{volume}{264}},
  \bibinfo{pages}{917} (\bibinfo{year}{1994}).

\bibitem[{\citenamefont{Brabec and Krausz}(2000)}]{cpa4}
\bibinfo{author}{\bibfnamefont{T.}~\bibnamefont{Brabec}} \bibnamefont{and}
  \bibinfo{author}{\bibfnamefont{F.}~\bibnamefont{Krausz}},
  \bibinfo{journal}{Rev.~Mod.~Phys.} \textbf{\bibinfo{volume}{72}},
  \bibinfo{pages}{545} (\bibinfo{year}{2000}).

\bibitem[{\citenamefont{McDermott et~al.}(1978)\citenamefont{McDermott,
  Marshall, Schlesinger, Parker, and Granatstein}}]{brs}
\bibinfo{author}{\bibfnamefont{D.~B.} \bibnamefont{McDermott}},
  \bibinfo{author}{\bibfnamefont{T.~C.} \bibnamefont{Marshall}},
  \bibinfo{author}{\bibfnamefont{S.~P.} \bibnamefont{Schlesinger}},
  \bibinfo{author}{\bibfnamefont{R.~K.} \bibnamefont{Parker}},
  \bibnamefont{and} \bibinfo{author}{\bibfnamefont{V.~L.}
  \bibnamefont{Granatstein}}, \bibinfo{journal}{Phys.~Rev.~Lett.}
  \textbf{\bibinfo{volume}{41}}, \bibinfo{pages}{1368} (\bibinfo{year}{1978}).

\bibitem[{\citenamefont{Esarey and Sprangle}(1992)}]{brs2}
\bibinfo{author}{\bibfnamefont{E.}~\bibnamefont{Esarey}} \bibnamefont{and}
  \bibinfo{author}{\bibfnamefont{P.}~\bibnamefont{Sprangle}},
  \bibinfo{journal}{Phys.~Rev.~A} \textbf{\bibinfo{volume}{45}},
  \bibinfo{pages}{5872} (\bibinfo{year}{1992}).

\bibitem[{\citenamefont{Sprangle and Drobot}(1979)}]{brs3}
\bibinfo{author}{\bibfnamefont{P.}~\bibnamefont{Sprangle}} \bibnamefont{and}
  \bibinfo{author}{\bibfnamefont{A.~T.} \bibnamefont{Drobot}},
  \bibinfo{journal}{Journal of Applied Physics} \textbf{\bibinfo{volume}{50}},
  \bibinfo{pages}{2652} (\bibinfo{year}{1979}).

\bibitem[{\citenamefont{Drake et~al.}(1974)\citenamefont{Drake, Kaw, Lee,
  Schmidt, Lis, and RosenBluth}}]{drake}
\bibinfo{author}{\bibfnamefont{J.~F.} \bibnamefont{Drake}},
  \bibinfo{author}{\bibfnamefont{P.~K.} \bibnamefont{Kaw}},
  \bibinfo{author}{\bibfnamefont{Y.~C.} \bibnamefont{Lee}},
  \bibinfo{author}{\bibfnamefont{G.}~\bibnamefont{Schmidt}},
  \bibinfo{author}{\bibfnamefont{C.~S.} \bibnamefont{Lis}}, \bibnamefont{and}
  \bibinfo{author}{\bibfnamefont{M.~N.} \bibnamefont{RosenBluth}},
  \bibinfo{journal}{Phys. Fluids} \textbf{\bibinfo{volume}{17}},
  \bibinfo{pages}{778} (\bibinfo{year}{1974}).

\bibitem[{\citenamefont{Tabak et~al.}(1994)\citenamefont{Tabak, Hammer,
  Glinsky, Kruerand, Wilks, Woodworth, Campbell, Perry, and Mason}}]{tabak}
\bibinfo{author}{\bibfnamefont{M.}~\bibnamefont{Tabak}},
  \bibinfo{author}{\bibfnamefont{J.}~\bibnamefont{Hammer}},
  \bibinfo{author}{\bibfnamefont{M.~E.} \bibnamefont{Glinsky}},
  \bibinfo{author}{\bibfnamefont{W.~L.} \bibnamefont{Kruerand}},
  \bibinfo{author}{\bibfnamefont{S.~C.} \bibnamefont{Wilks}},
  \bibinfo{author}{\bibfnamefont{J.}~\bibnamefont{Woodworth}},
  \bibinfo{author}{\bibfnamefont{E.~M.} \bibnamefont{Campbell}},
  \bibinfo{author}{\bibfnamefont{M.~J.} \bibnamefont{Perry}}, \bibnamefont{and}
  \bibinfo{author}{\bibfnamefont{R.~J.} \bibnamefont{Mason}},
  \bibinfo{journal}{Physics of Plasmas} \textbf{\bibinfo{volume}{1}},
  \bibinfo{pages}{1626} (\bibinfo{year}{1994}).

\bibitem[{\citenamefont{Son and Fisch}(2005{\natexlab{a}})}]{sonprl}
\bibinfo{author}{\bibfnamefont{S.}~\bibnamefont{Son}} \bibnamefont{and}
  \bibinfo{author}{\bibfnamefont{N.~J.} \bibnamefont{Fisch}},
  \bibinfo{journal}{Phys.~Rev.~Lett.} \textbf{\bibinfo{volume}{95}},
  \bibinfo{pages}{225002} (\bibinfo{year}{2005}{\natexlab{a}}).

\bibitem[{\citenamefont{Son and Fisch}(2004)}]{sonpla}
\bibinfo{author}{\bibfnamefont{S.}~\bibnamefont{Son}} \bibnamefont{and}
  \bibinfo{author}{\bibfnamefont{N.~J.} \bibnamefont{Fisch}},
  \bibinfo{journal}{Phys.~Lett.~A} \textbf{\bibinfo{volume}{329}},
  \bibinfo{pages}{16} (\bibinfo{year}{2004}).

\bibitem[{\citenamefont{Son and Fisch}(2006{\natexlab{a}})}]{sonpla2}
\bibinfo{author}{\bibfnamefont{S.}~\bibnamefont{Son}} \bibnamefont{and}
  \bibinfo{author}{\bibfnamefont{N.~J.} \bibnamefont{Fisch}},
  \bibinfo{journal}{Phys.~Lett.~A} \textbf{\bibinfo{volume}{356}},
  \bibinfo{pages}{65} (\bibinfo{year}{2006}{\natexlab{a}}).

\bibitem[{\citenamefont{Son and Fisch}(2006{\natexlab{b}})}]{sonpla3}
\bibinfo{author}{\bibfnamefont{S.}~\bibnamefont{Son}} \bibnamefont{and}
  \bibinfo{author}{\bibfnamefont{N.~J.} \bibnamefont{Fisch}},
  \bibinfo{journal}{Phys.~Lett.~A} \textbf{\bibinfo{volume}{356}},
  \bibinfo{pages}{72} (\bibinfo{year}{2006}{\natexlab{b}}).

\bibitem[{\citenamefont{Son and Fisch}(2005{\natexlab{b}})}]{sonchain}
\bibinfo{author}{\bibfnamefont{S.}~\bibnamefont{Son}} \bibnamefont{and}
  \bibinfo{author}{\bibfnamefont{N.~J.} \bibnamefont{Fisch}},
  \bibinfo{journal}{Phys.~Lett.~A} \textbf{\bibinfo{volume}{337}},
  \bibinfo{pages}{397} (\bibinfo{year}{2005}{\natexlab{b}}).

\bibitem[{\citenamefont{McKinstrie and Simon}(1986)}]{McKinstrie}
\bibinfo{author}{\bibfnamefont{C.~J.} \bibnamefont{McKinstrie}}
  \bibnamefont{and} \bibinfo{author}{\bibfnamefont{A.}~\bibnamefont{Simon}},
  \bibinfo{journal}{Phys.~Fluids} \textbf{\bibinfo{volume}{29}},
  \bibinfo{pages}{1959} (\bibinfo{year}{1986}).

\bibitem[{\citenamefont{Son}(2010)}]{sonbrs}
\bibinfo{author}{\bibfnamefont{S.}~\bibnamefont{Son}}, \bibinfo{journal}{arixv}
  \textbf{\bibinfo{volume}{0}}, \bibinfo{pages}{0} (\bibinfo{year}{2010}).

\bibitem[{\citenamefont{Son et~al.}(2010)\citenamefont{Son, Ku, and
  Moon}}]{sonbackward}
\bibinfo{author}{\bibfnamefont{S.}~\bibnamefont{Son}},
  \bibinfo{author}{\bibfnamefont{S.}~\bibnamefont{Ku}}, \bibnamefont{and}
  \bibinfo{author}{\bibfnamefont{S.~J.} \bibnamefont{Moon}},
  \bibinfo{journal}{Phys.~Plasmas} \textbf{\bibinfo{volume}{17}},
  \bibinfo{pages}{114506} (\bibinfo{year}{2010}).

\end{thebibliography}

\end{document}